\documentclass[a4paper]{article}
\usepackage{amssymb}

\title{Quantum Inflation of Classical Shapes}
\author{Tim Koslowski\\ Department of Mathematics and Statistics\\University of New Brunswick\\Fredericton, NB, Canada\\ \texttt{t.a.koslowski@gmail.com}}

\begin{document}

\maketitle

\begin{abstract}
  \noindent I consider a quantum system that possesses key features of quantum shape dynamics and show that the evolution of wave-packets will become increasingly classical at late times and tend to evolve more and more like an expanding classical system. At early times however, semiclassical effects become large and lead to an exponential mismatch of the apparent scale as compared to the expected classical evolution of the scale degree of freedom. This quantum inflation of an emergent and effectively classical system, occurs naturally in the quantum shape dynamics description of the system, while it is unclear whether and how it might arise in a constrained Hamiltonian quantization.
\end{abstract}

\section{Introduction}

Most approaches to quantum gravity focus on finding a quantum theory of spacetime geometry. This is natural, because classical General Relativity describes gravity as a geometry of spacetime that satisfies Einstein's equations. However, since so far no satisfactory theory of quantum gravity has been produced, one is lead to ask whether classical gravity can be described by a different set of principles and whether the implementation of these principles can provide guidance in the search for quantum gravity. Shape dynamics is such an alternative description in which classical gravity is described as the evolution of spatial conformal geometry \cite{Anderson:2004bq,Barbour:2011dn,Barbour:2010dp,Gomes:2010fh}. The predictions of classical shape dynamics are nevertheless locally indistinguishable from classical General Relativity \cite{Gomes:2011zi,Gomes:2011au,Gomes:2012hq,Koslowski:2012uk}. 

The hope of the shape dynamics program is that this reformulation of classical gravity will reveal new strategies to quantizing gravity \cite{Koslowski:2013gua}. The resulting quantum theory might then differ significantly from a quantum theory of spacetime. This leads to the question: ``What differences can we expect from quantum shape dynamics as compared to quantum spacetime geometry?'' This is a very reasonable question, since two quantum theories can, when constructed from different sets of principles, lead to completely different quantum theories.\footnote{For example Wheeler-DeWitt quantum cosmology and loop quantum cosmology make significantly different predictions, which can be traced back to the fact that Wheeler-DeWitt quantum cosmology is based on canonical commutation relations, while loop quantum cosmology is, loosely speaking, based lattice gauge theory \cite{Bojowald:2008zzb}.}

Since there is no quantum theory of full shape dynamics, I will consider a simple quantum toy model whose classical analogue admits two formulations: one formulation possesses key features of the spacetime description of gravity and the other shares key features with shape synamics. I then compare the quantum theories associated with these two formulations. It is instructive to look at the relation of General Relativity and shape dynamics to understand what these key features are. I start with the ADM formulation  \cite{Arnowitt:1962hi} of General Relativity.

The fundamental degrees of freedom of the ADM formulation are the spatial metric $g_{ab}$ and its canonically conjugate momentum density $\pi^{ab}$. The ADM-Hamiltonian $\mathbf{H}=S(N)+H(\xi)$ is a linear combination\footnote{I use the smearing notation, so e.g. $H(\xi)=\int H_a(x) \xi^a(x)$.} of gauge generators: the vector constraint $H_a(x)$, which generates spatial diffeomorphisms, and the scalar constraint $S(x)$, whose action entangles refoliations (which are pure gauge) with time reparametrizations (i.e. gravitational dynamics). Shape Dynamics uses the same phase space as the ADM formulation, but disentangles gravitational dynamics from the gauge problem by replacing the scalar ADM constraints with the generators $D(x)$ of spatial conformal transformations and a physical Hamiltonian $H_{SD}(\tau)$ that depends explicitly on York time $\tau$. 

The shape dynamics formulation of classical gravity is nevertheless locally indistinguishable from General Relativity. The simplest way to explain this uses the fact that Shape Dynamics can be obtained as gauge-unfixing General Relativity in York gauge. This process interchanges the role of gauge generators and gauge fixing conditions. The gauge fixing conditions for York gauge turn out to be the generators of spatial conformal transformations\footnote{This is subtle procedure: The gauge fixing condition for York gauge is the generator of conformal transformations that do not change the total spatial volume, while the generator of total volume changes turns out to be a time variable (York time). The trading of the ADM scalar constraints for conformal transformations is thus a two step process that involves the trading of the local scalar constraints and a subsequent deparametrization w.r.t. York time.} \cite{Dirac:1958ha,Dirac:1959fi,York:1972ro,York:1973cm}. Local scale is thus, form the perspective of 
shape dynamics, pure gauge. The equivalence with General Relativity however brings an apparent scale into play. This apparent scale can be understood in two ways:
\begin{enumerate}
 \item {\it Formally:} The equivalence of shape dynamics is manifest with the ADM formulation, i.e. the trajectories coincide completely and not just up to pure gauge, if the scalar ADM constraints are used as a gauge-fixing for the spatial Weyl constraints. This gauge-choice of scale allows one to ``see'' the Einsteinian spacetime in a shape dynamics trajectory \cite{Gomes:2012hq}.
 \item {\it Operationally:} If standard matter fields are coupled to shape dynamics, then the spacetime geometry can be deduced from the evolution of ``test'' fluctuations of matter fields. The apparent scale deduced from the evolution of matter fields coincides with the apparent scale that has been deduced using the formal method \cite{Gomes:2011au}. This is why I will use the formal method in this paper and reserve the operational method for forthcoming work.
\end{enumerate}
The key feature of the duality between shape dynamics and General Relativity is that the Hamilton constraints of General Relativity are traded for Weyl-constraints and a physical Hamiltonian of shape dynamics through a mechanism that does not change any physical predictions. 

In section \ref{sec:ToyModel} I will introduce a class of particle models\footnote{Similar particle models have proven to provide very useful guidance in the development of relational dynamics, which has lead to the shape dynamics formulation of gravity \cite{Anderson:2010xm,Barbour:1982gha,Barbour:2013qb,Barbour:2013goa}.} that admit the same kind of duality. The two classical descriptions are then quantized in sections \ref{sec:ConstrainedHamiltonian} and \ref{sec:QuantumSD}. In this process I pay particular attention to semiclassical equations of motion\footnote{There are many ways to obtain effective semiclassical equations of motion. In this paper I use approximations to Bohmian equations of motion \cite{Bohm:1952aa} for a configuration at the peak of a semiclassical wave packet. This method is explained in the appendices and is related to Bohmian mechanics. I caution that determining the accuracy of this method of taking a semiclassical limit is a delicate issue, and the present paper represents only a 
first step in which sufficient accuracy is assumed.  A more detailed discussion of the classical limit in Bohmian mechanics can e.g. be found in \cite{Allori:2002dg}.}. 

This analysis shows that the quantum shape dynamics formulation of the toy model has a natural mechanism by which the classical approximation to the evolution of a semiclassical wave packet becomes better and better at late times. This is a prerequisite for the ``geometrogenesis'' of ``hylogenesis'' process proposed in \cite{Barbour:2013jya}, i.e. the generation of an apparent classical spacetime from quantum shape dynamics. The mechanism behind this is the fact that the shape dynamics analogue of what is called ``quantum potential'' in Bohmian mechanics decays with time. A second consequence of the shape dynamics analogue of the quantum potential turns out lead to an exponentially large correction to the apparent scale at early times (when $|\tau|$ is small). This inflation of scale is not produced by the constrained Hamiltonian formulation. This suggests, by analogy with quantum cosmology of shape dynamics, that inflation could have a natural explanation as semiclassical effects of quantum shape dynamics. 
Forthcoming work will be concerned with the investigation of this question.

\section{A Class of Shape Dynamics Toy Models}\label{sec:ToyModel}

I will now consider a very simple class of particle models that possesses the key features of the duality between shape dynamics and General Relativity. I.e. I consider models that can on the one hand be described as a constrained Hamiltonian system (General Relativity description) and on the other hand as a system with time-dependent physical Hamiltonian in which scale is pure gauge (the analogue of the shape dynamics description). For this, I consider a particle with configuration $q^a\,:\,a=1,2$ and canonically conjugate momenta $p_a\,:\,a=1,2$. Moreover, I assume a standard Hamiltonian of the form $H=\frac{p^2}{2m}+V(q)$, where the potential $V(q)<0$ is homogeneous of degree $k>-2$ in the $q^a$'s. If the system is isolated, the initial energy $E$ of this system will be conserved by the evolution and the initial energy and evolution can be jointly encoded in a Hamilton constraint
\begin{equation}
 \chi_E=\frac{p^2}{2m}+V(q)-E \approx 0,
\end{equation}
which generates reparametrizations of time. The closest analogy with pure gravity is attained for $E=0$, which is the case that I will consider form now on. It is convenient to analyze this system after performing a canonical transformation to polar coordinates $(r,\phi)$ under which the zero-energy Hamilton constraint $\chi$ changes to
\begin{equation}\label{equ:HamiltonConstraint}
 \chi=\frac{p_r^2+r^{-2}p_\phi^2}{2m}-r^k\,C_{sh}(\phi)\approx 0,
\end{equation}
where I defined the shape complexity $C_{sh}(\phi):=-r^{-k}\,V(r,\phi)>0$ (see \cite{Barbour:2013jya} for properties of the shape complexity). It follows that the forward evolution of $\tau:=r\,p_r$ is monotonic, since
\begin{equation}
 \dot \tau = \{\tau,\chi\}\approx (k+2) r^k C_{sh}(\phi)\ge 0.
\end{equation}
We can thus use $\tau$ as a time variable and deparametrize the system. This is simplified by the canonical transformation $(r,p_r) \to (\lambda=\textrm{ln}(r),\tau=p_r\,r)$:
\begin{equation}
 r^2\chi=\frac{\tau^2+p_\phi^2}{2m}-e^{(k+2)\lambda}\,C_{sh}(\phi)\approx 0,
\end{equation}
which allows us to find the shape Hamiltonian $H_{sh}$ by solving $\chi=0$ for $\lambda$, the variable canonically conjugate to $\tau$:
\begin{equation}\label{equ:physicalHamiltonian}
 H_{sh}=\frac{1}{k+2} \ln\left(\frac{\tau^2+p_\phi^2}{2m\,C_{sh}(\phi)}\right),
\end{equation}
where $\tau$ is not a dynamical variable, but the time-parameter for the evolution of the shape degree of freedom $\phi$. The shape dynamics description produces exactly the same dynamics of $\phi$ as a Newtonian system with Hamiltonian $H=\frac{p^2}{2m}+V(q)$ and energy $E$ does. The manifest equivalence between the two descriptions can be seen, if we ``Newtonianize'' a shape dynamics trajectory $\phi(\tau)$ with an {\bf apparent} scale $r(\tau)=e^{H_{sh}(\tau)}$. In other words: It is this apparent scale that one would deduce from observations within the system (i.e. without an external rod) under the assumption of a standard Newtonian description.

The equivalence of the two descriptions allows us to describe the same physics once with a Hamilton constraint (\ref{equ:HamiltonConstraint}) with scale $r$ and once as a scale-invariant system with physical Hamiltonian (\ref{equ:physicalHamiltonian}). The most natural Schr\"odinger quantization of the Hamilton constraint is to replace $p^2 \to -\hbar^2 \Delta$, leading to the Hamilton constraint (or ``Wheeler-DeWitt'') equation
\begin{equation}\label{equ:QuantumHamiltonConstraint}
 \hat \chi \Psi(q)= -\frac{\hbar^2}{2m} \Delta \Psi(q) +V(q)\Psi(q)=0.
\end{equation}
A similarly natural Schr\"odinger quantization for the physical Hamiltonian is to solve the operator ordering ambiguity by writing the logarithm of the fraction as a difference of logarithms\footnote{To simplify notation I omit constant factors that are needed to render the arguments of the logarithms dimensionless. This is permissible because the subsequent discussion is independent of these factors.} and to make the replacement $p_\phi \to -i\hbar \partial_\phi$, which leads to the Schr\"odinger equation
\begin{equation}\label{equ:Schroedinger}
 i\hbar \partial_\tau \Psi(q)=\frac{\ln\left(\tau^2-\hbar^2 \partial^2_\phi\right)}{k+2}-\frac{\ln\left(\,C_{sh}(\phi)\right)}{k+2},
\end{equation}
where the RHS is self-adjoint on the span of $\Psi_n(\phi)=e^{in\phi}$. In the following, I will compare the semiclassical behavior of these two quantizations.

\section{Quantum Hamilton Constraint}\label{sec:ConstrainedHamiltonian}

The standard approach to quantum cosmology is to assume that the wave function of the universe satisfies a Wheeler-DeWitt equation (see e.g. \cite{Rovelli:2004tv} for a general introduction and \cite{Bojowald:2010xp} for a specific proposal). Let us now use the analogous approach in the particle model and assume that the wave function $\psi$ of the particle system solves the Hamilton constraint equation (\ref{equ:QuantumHamiltonConstraint}). We would now like to understand the semiclassical limit of $\psi$ to be able to compare it with the analogue of the shape dynamics quantization in the next section. 

Unfortunately, the timeless nature of the Hamilton constraint equation precludes the direct application of the Bohmian trajectory method\footnote{See appendix \ref{appendix:Trajectory} for a description of this method.} for the investigation of the semiclassical limit, as we will use it in the next section. However, we can still proceed in close analogy with this approach by inserting $\psi=R\,e^{\frac i \hbar S}$ in the constraint equation (\ref{equ:QuantumHamiltonConstraint}) and separate the real and imaginary part, which implies
\begin{equation}\label{equ:WdWsplit}
 \begin{array}{rcl}
   0&=&(S_{,r})^2+r^{-2}(S_{,\phi})^2-2m\left(r^k\,C_{sh}(\phi)+\hbar^2\,\frac{R_{,rr}+r^{-2}R_{,\phi\phi}}{R}\right)\\
   0&=&\partial_r\left(R^2\,S_{,r}\right)+\partial_\phi\left(R^2\,r^{-2}S_{,\phi}\right).
 \end{array}
\end{equation}
The first line is, in analogy with e.g. a free relativistic particle $g^{\mu\nu}S_{,\mu}S_{,\nu}-m^2=0$, a relativistic Hamilton-Jacobi equation
\begin{equation}
  g^{\mu\nu}S_{,\mu}S_{,\nu}-V\,=\,0, 
\end{equation}
which is modified by adding $\hbar^2$ times the quantum potential $Q=-\frac{g^{\mu\nu}R_{,\mu\nu}}{R}$. The second equation is, in analogy with $\partial_\mu\left(\rho\,u^\mu\right)=0$, a relativistic continuity equation from which we find the relativistic velocity
\begin{equation}
 u^\mu=\left(\,S_{,r}\,\,,\,\,r^{-2} S_{,\phi}\,\right)=g^{\mu\nu} S_{,\nu}.
\end{equation}
One should, in analogy with the relativistic particle, interpret these constraint equations as evolution equations in a ``physical clock'' variable. The physical clock variable that corresponds classically to the shape dynamics evolution is $\tau=p_r\,r$, which is related to the (locally monotonic) clock variable $t:=r$ that I will use now. The first line of equation (\ref{equ:WdWsplit}) then turns into the standard Hamilton-Jacobi equation
\begin{equation}\label{equ:WdWHJ}
 0=\dot S\,\pm\,\sqrt{2m\,\left(r^k C_{sh}(\phi)+\hbar^2\,Q\right)-\left(\frac{S^\prime}{r}\right)^2},
\end{equation}
where I used a dot to denote $t$-derivatives. The sign $\pm$ reminds us that $r$ is only locally a good clock, because the same radius is attained during once during the contracting phase (when $p_r<0$) and once during the expanding phase (when $p_r>0$). Moreover, $r$ is not a good physical clock in the neighborhood of $p_r=0$. An $\hbar$ expansion of equation (\ref{equ:WdWHJ}) gives the Hamiltonian $H^r_{cl}$ for the classical $r$-evolution
\begin{equation}
 \begin{array}{rcl}
   H^r_{cl}&=&\sqrt{2m\,r^k\,C_{sh}(\phi)-\left(\frac{p_\phi}{r}\right)^2}.
 \end{array}  
\end{equation}
We thus come to the expected and unspectacular conclusion that a wave packet can be expected to evolve classically in $r$ when the effect of $\delta H_{qu}$ on the constraint equation is negligible. 

The semiclassical limit presented in this section depends critically on the choice of internal clock, which can be chosen arbitrarily. The discussion presented in this section should therefore be understood as an attempt to use techniques of standard quantum cosmology (as e.g. proposed in \cite{Bojowald:2010xp}) in the investigation of the toy model, in order to be able to compare this with the shape dynamics formulation in the next section. The approach suffers from several weaknesses; the foremost is the arbitrariness in the choice of internal clock. Furthermore:
\begin{enumerate}
 \item Even when a preferred internal clock is chosen, reinterpreting the quantum Hamilton constraint as an evolution equation in this variable does in general not lead to unitary evolution. 
 \item Quantum effects can be arbitrarily large, even if the reduced wave function $\psi_r$ at a given time $r=t_o$, i.e. $\psi_r(\phi):=\psi(r,\phi)$, is an extremely semiclassical wave packet. This is due to the contribution of $R_{,rr}$ to the quantum potential which is not constrained by the assumption that the reduced wave function is semiclassical.
\end{enumerate}
These two points make the procedure of turning a quantum degree of freedom into a classical clock variable a very delicate procedure. Generally one can only expect that the mechanism is reliable in a suitable semiclassical limit. However, a suitably semiclassical wave function has to be assumed by hand, because the quantum system does not provide a mechanism that makes the evolution increasingly classical.

\section{Quantum Dynamics of Shapes}\label{sec:QuantumSD}

Quantum Shape Dynamics describes a closed quantum system as Schr\"odinger evolution on shape space and thus evades the conceptual problems of the quantum constraint system on configuration space. One can thus directly apply the semiclassical approach described in appendixes \ref{appendix:Trajectory} and \ref{appendix:Nonstandard}. This procedure starts from the classical Hamiltonian
\begin{equation}\label{equ:classSDhamiltonian}
 H^{cl}=\frac{1}{k+2}\left(\ln\left(\frac{\tau^2+p_\phi^2}{\tau_o^2}\right)-\ln\left(\frac{C_{sh}(\phi)}{C_o}\right)\right),
\end{equation}
where $\tau_o$ and $C_o$ are constants that make the arguments of the logarithms dimensionless, but whose values do not influence the classical evolution and only appear as an unobservable overall phase in the quantum evolution. This Hamiltonian can be quantized as in equation (\ref{equ:Schroedinger}).

The leading order Hamiltonian in an $\hbar$-expansion is according to appendix \ref{appendix:Nonstandard} given by $H^{cl}$ of equation (\ref{equ:classSDhamiltonian}). The subleading order in the $\hbar$-expansion is according to equation (\ref{equ:nonStandardHJ}) given by
\begin{equation}
 \begin{array}{l}
   \delta H=-\frac{\hbar^2}{k+2}\left(\frac{R^{(2)}(\phi)}{2\,R(\phi)}\frac{p_\phi^2-\tau^2}{(p_\phi^2+\tau^2)^2}+2 \left(\frac{R^\prime(\phi)S^{(2)}(\phi)}{R(\phi)}+\frac 1 3 S^{(3)}(\phi)\right)\frac{p_\phi^3+3p_\phi \tau^3}{(p_\phi^2+\tau^2)^3}\right.\\
   \quad\quad\quad\quad\quad\quad\left.+\frac 3 2 \left(S^{(2)}\right)^2\frac{6p_\phi^2\tau^2-(p_\phi^4+\tau^4)}{(p_\phi^2+\tau^2)^4}\right)
 \end{array}
\end{equation} 
Let us now consider the effective equations of motion: the leading order in the $\hbar$ expansion are, according to appendix \ref{appendix:Nonstandard}, given by the equations of motion generated by $H^{cl}$. The subleading contributions to the effective equations of motion are, as explained in appendix \ref{appendix:EffectiveEOM}, not the Hamiltonian equations of motion derived from $\delta H$. Rather, the subleading order of the equation for $\phi$ is given by the subleading order of the velocity equation (\ref{equ:nonStandardVelocity}) and the subleading order to $p_\phi$ is, according to equation (\ref{equ:dotPsubleading}), given by $2 \hbar^2 s_2\,v_{qu}(q)-\left.(\delta H)_{,x}\right|_{x=q,s_1=p}$. 

The $p_\phi$-dependence of $H^{cl}$ implies that $\frac{\partial^n H^{cl}}{\partial p_\phi^n} = \mathcal O(\tau^{-n})$ for even $n$ and $\mathcal O(\tau^{-(n+1)})$ for odd $n$ in the limit of large $\tau$. Thus, $\delta H$ becomes negligible for large $\tau$, which implies that the system provides a natural mechanism by which the classical evolution equations for a semiclassical wave packet become dominant at late times. This is a consequence of the time-dependence of the non-standard kinetic term of the Hamiltonian (\ref{equ:Schroedinger}). 

This decay of quantum corrections provides an ex post facto justification of use of the Bohmian trajectory method, because it implies that $(\phi,p_\phi)$ decouple effectively from the rest of the wave function (i.e. the $R,\tilde S$ of appendix \ref{appendix:Trajectory}). Moreover, if a wave function $\psi_\tau(\phi)$ is semiclassical, then the Schr\"odinger evolution is, by assumption, semiclassical. Notice that this seemingly tautological statement is not true for the constrained Hamiltonian quantization, as we discussed at the end of the preceding section.

\section{Unexpectedly large Quantum Effects}

The semiclassical effects derived in the previous two sections are most interesting in the quantum to classical transition regime. I.e. when the classical terms already dominate the equations of motion, and provide a reasonable approximation, but the where at the same time the effect of the leading quantum corrections is not negligible. The precise determination of this regime is a subtle issue, since small deviations in the equations of motion can, e.g. through resonance effects, lead to large integrated effects. A precise discussion of integrated effects is beyond the scope of this paper; rather I will focus on a different mechanism that has the potential to generate large semiclassical effects.

Recall that shape dynamics implements the relational principle that the overall scale a closed system (a model universe) is unobservable. This means practically that the variable $r$ in the toy model of section \ref{sec:ToyModel} is not observable. This apparent scale degree of freedom can however be deduced from the assumption that certain reference subsystems $l_o$ of the universe provide an unchanging unit rod, such that the relational quantity $r/l_o$ emerges as the apparent scale of the universe. If the assumption that the size of a certain reference system is constant is based on classical energy conservation, analogous to equation (\ref{equ:HamiltonConstraint}), then the apparent scale will be derived from the assumption that the energy conservation constraint holds. Consequently, the classical scale derived from (\ref{equ:physicalHamiltonian}) is
\begin{equation}\label{equ:apparentScale}
 r_{cl}=l_o\,\exp\left(H_{sh}\right).
\end{equation}
Thus, if the quantum corrections to the equations of motion of the configuration are small, so the quantum shape dynamics evolution of a peaked wave packet is well approximated its associated classical velocity, the apparent scale can still experience exponentially large quantum effects, due to quantum corrections to the effective Hamiltonian.

Let us investigate this possibility in the present toy model: First of all, we discussed in the previous section that the quantum shape dynamics evolution of a wave packet becomes increasingly semiclassical as $\tau$ increases, which tells us that we have to look at the late time $\tau$-dependence to observe the quantum to classical transition. We then use the fact that the quantum corrections to the velocity equations depend on $\frac{\partial^n H^{cl}}{\partial p_\phi^n}$ with $n\ge 4$, such that the quantum corrections to the peak evolution decay as $\tau^{-4}$ as $\tau$ increases, while $\delta H$ decays as $\tau^{-2}$ as $\tau$ increases. Thus, the evolution of the configuration becomes classical much faster than the $\delta H$, which is the logarithm of the apparent scale. We thus see that the evolution of the peak of a wave packet will become $\tau^2$ times faster classical than the evolution of the momenta and the derived logarithm of the apparent scale. However, when $|\tau|$ is small, we find a non-
negligible $\delta H$, which implies that the apparent scale will mismatch the classical predictions by the exponential factor $e^{\delta H}$. 
\begin{equation}
 r=l_o\,\exp\left(H_{sh}+\delta H\right)=r_{cl}\,e^{\delta H}
\end{equation}
The quantum shape dynamics evolution thus provides a mechanism for exponentially large (or ``inflationary'') deviations in a regime where the peak of the wave packet appears to evolve according to a classical velocity law.

This effect, i.e. the effect that the peak evolution of a wave packet follows a classical law like equation (\ref{equ:Guidance}) while the quantum corrections $-\frac{\hbar^2}{2m}\frac{\Delta R}{R}$ to the Hamiltonian may not be negligible, is well known in Bohmian mechanics as the effect of the quantum potential and is not special to shape dynamics. What is however special to shape dynamics is the fact that the shape dynamics analogue of the quantum potential, i.e. $\delta H$, enters the apparent scale as an exponential correction factor $e^{\delta H}$, so the apparent scale is inflated by quantum corrections. 

\section{Discussion}

In this paper I considered the quantization of a class of simple models that shares key features with the classical duality between the shape dynamics description of gravity and the description of gravity through General Relativity. The analogue of General Relativity is a system with constrained Hamiltonian. The quantization of this formulation leads to a timeless quantum constraint equation, analogous to the Wheeler-DeWitt equation. I interpret this system in the usual way, by interpreting one of its degrees of freedom as an internal clock, so the constrained wave function evolves w.r.t. this clock. This leads to an expected semiclassical limit. Since the dynamics of the system is only an effective dynamics w.r.t. an arbitrarily chosen internal clock, one can not say much more about quantum effects on the system's dynamics, since all such statements depend on details of the (arbitrarily chosen) internal clock.

The analogue of the shape dynamics description is a description of the system in which the scale degree of freedom has been eliminated and which evolves w.r.t. a time-dependent physical Hamiltonian. The quantization of this system leads to an analogue of Moncrief's quantization of pure gravity in 2+1 dimesnions \cite{Moncrief:1989dx,Budd:2011er}. The leading order in an $\hbar$-expansion produces the expected classical limit. It is sensible to investigate the leading order quantum effects on the dynamics, since the system possesses an unambiguous dynamics, i.e. its wave function satisfies a Schr\"odinger equation rather than a constraint equation. A reader familiar with Bohmian mechanics will now expect effects due to ``contributions of the quantum potential.'' These expected effects appear in the shape dynamics formulation, but the shape dynamics analogue of the quantum potential has unexpected features. These new features can be traced to features of shape dynamics that do not occur in standard quantum 
systems. These are:
\begin{enumerate}
 \item The shape dynamics Hamiltonian possesses an explicit time-dependence and the Hamiltonian is not quadratic in momenta. This turns out to provide a mechanism for late time classicality, because the analogue of the quantum potential turns out to decay with time. This seems to make the present semiclassical analysis very stable and may turn out to be sufficient to initiate the ``geometrogenesis'' mechanism that was proposed in \cite{Barbour:2013jya}.
 \item Scale is not a dynamical degree of freedom in shape dynamics; rather scale appears as a derived auxiliary concept. This derived ``apparent scale'' receives contributions from the shape dynamics analogue of the quantum potential and it turns out that these lead to exponentially large corrections that ``inflate'' the apparent scale at early times, i.e. when $|\tau|$ is small.
\end{enumerate}
These two features of the particle model could have important consequences for quantum cosmology based on shape dynamics. They suggest, if the analogy with the toy model holds, that shape dynamics may aid the emergence of classical physics at late times and moreover provide an explanation for inflation.

I conclude this paper with a caution: This paper represents only a first look at the possible consequences of trading the traditional formulation of isolated quantum systems for a quantum version of shape dynamics. The present investigation of the toy model should therefore be understood as a motivation for detailed research of similar effects on the shape dynamics formulation of gravity, in particular in mini-superspace models and models in which the apparent scale is deduced operationally. Such models are currently under investigation. It would of course be truly remarkable if some of these  quantum effects would be found in the shape dynamics approach to quantum cosmology, as these could have important consequences for cosmology. E.g. if the apparent scale receives exponentially large quantum corrections through shape dynamics then cosmological inflation could be driven by the shape dynamics analogue of the quantum potential.

\subsection*{Acknowledgements}

This work was supported in part by the Natural Sciences and Engineering Research Council of Canada (NSERC) through a grant to the University of New Brunswick and by the Foundational Questions Institute through grant FQXi-RFP3-1339. I am very grateful for an invitation to the 2013 ``Haunted Workshop'' in Tepoztlan, Mexico, where discussions with Ward Struyve and Daniel Sudarsky raised the question about quantum corrections to classical shape dynamics cosmology was raised. It was also at this workshop where Ward Struyve introduced me to detail about Bohmian mechanics.

\begin{appendix}

\section{Bohmian Trajectory Approach to Classical Limit}\label{appendix:Trajectory}

The Bohmian trajectory method is a useful tool in the investigation of classical limits and allows one to quantify the deviations from classicality. This methods can be summarized as follows: One starts with Bohm's formulation of quantum mechanics \cite{Bohm:1952aa}. This formulation uses, in addition to the quantum mechanical wave function $\psi$ that evolves according to the Schr\"odinger equation $i\hbar\,\partial_t\,\psi=\hat H \,\psi$, a configuration $Q^a$ to a physical system which evolves according to the guidance equation $ \partial_t Q^a = \frac{j_\psi^a(Q)}{|\psi|^2(Q)}$ (where $j_\psi^a(x)$ denotes the quantum mechanical probability current associated with the wave function $\psi$). It follows that an initial $|\psi|^2$-distributed particle density that follows the guidance equation will remain in this so-called quantum equilibrium distribution. Thus, the peak of a wave packet will approximately evolve like the Bohmian configuration $Q^a$. The wave function $\psi(x)$ is, in this picture, a 
generalization of the classical momenta $P_a$. The classical limit is attained when the evolution of $Q^a$ is governed by approximately classical equations, i.e. all degrees of freedom of $\psi$, except for $P_a:=-i \partial_a \textrm{arg}(\psi(x))|_{x^a=Q^a}$ decouple approximately \cite{Allori:2002dg}, so the the evolution equations of $(Q^a,P_b)$ turn to Hamilton's equations plus negligible contributions from the remaining generalized momenta present in $\psi$. This decoupling limit can be systematically investigated by inserting the ansatz $\psi(x)=R(x)\,e^{\frac i h S(x)}$ in Schr\"odinger's equation, which gives for standard Hamiltonians $H=\frac{p^2}{2m}+V(q)$:
\begin{equation}\label{equ:SCsplit}
 \begin{array}{rcl}
   -\partial_t R^2 &=& \textrm{div}\left(\frac{\nabla S}{m} \,R^2\right)\\
   -\partial_t S   &=& \frac{(\nabla S)^2}{2m}+V-\frac{\hbar^2}{2m}\frac{\Delta R}{R}
 \end{array}
\end{equation}
and the guidance equation becomes
\begin{equation}\label{equ:Guidance}
   \partial_t Q^a = \left.\frac{\nabla^a S(x)}{m}\right|_{x\equiv Q}.
\end{equation}
Let us set $S(x):=P_a(Q^a-x^a)+\tilde S(x)$, with $P^a=\nabla^a S(x)|_{x\equiv Q}$. It follows from the fact that the second line of equation (\ref{equ:SCsplit}) is a Hamilton-Jacobi equation with additional quantum potential $V_{qu}(Q)=-\frac{\hbar^2}{2m}\frac{\Delta R}{R}(Q)$ that $P_a$ evolves according to the Hamilton 
\begin{equation}
 \partial_t P_a = \{P_a,H_{qu}\},
\end{equation}
with the effective Hamiltonian $H_{qu}(P,Q)=\frac{P^2}{2m}+V(Q)+V_{qu}(Q)$, which happens to be the same Hamiltonian that generates the guidance equation (\ref{equ:Guidance}) as
\begin{equation}
 \partial_t Q^a = \{Q^a,H_{qu}(P,Q)\}.
\end{equation}
We see that, for a standard Hamiltonian, the only coupling of the equations of motion for $(Q^a,P_b)$ with $(R(x),\tilde S(x))$ is through the quantum potential $V_{qu}(Q)$. Hence, $(Q^a,P_b)$ evolve approximately classically if the quantum force $F_b^{qu}\frac{1}{m}\{P_b,V_{qu}(Q)\}$ is negligible. Similarly, we expect that the leading semiclassical effects can be effectively understood in terms of the quantum force.

\section{Nonstandard Hamiltonians}\label{appendix:Nonstandard}

The idea behind the Bohmian trajectory approach to the classical limit can be applied to nonstandard Hamiltonians, but the equations analogous to (\ref{equ:SCsplit}) can only be calculated as a series in $\hbar$. Motivated by the form the Shape Dynamics Hamiltonian, let us consider a Hamiltonian of the form
\begin{equation}\label{equ:nonStandardHamiltonianForm}
 \hat H = f(-i\hbar \partial) + g(q).
\end{equation}
Using the semiclassical ansatz $\psi=R\,e^{\frac i \hbar S}$ and the commutation relation 
\begin{equation}
 \begin{array}{ll}
   &e^{-\frac i \hbar S}f\left(-i\hbar \,\partial \right)\,R\,e^{\frac i \hbar S}\\ 
   =&R\,f(S^\prime) - i\hbar\left(R^\prime f^\prime(S^\prime)+\frac 1 2 R S^{(2)}f^{(2)}(S^\prime)\right)\\
   -&\hbar^2\left(\frac 1 2 R^{(2)} f^{(2)}(S^\prime)+\left(\frac 1 2 R^\prime S^{(2)}+\frac 1 6 R S^{(3)}\right) f^{(3)}(S^\prime)+\frac 1 8 R (S^{(2)})^2 f^{(4)}(S^\prime)\right)\\
   +&i\,\hbar^3\left(\left(\frac 1 4 R^{(2)}S^{(2)}+\frac 1{4!}R S^{(4)}\right) f^{(4)}(S^\prime)+\left(\frac 1 8 R^\prime (S^{(2)})^2+\frac 1{12} R S^{(2)}S^{(3)}\right)f^{(5)}(S^\prime)\right.\\
   &\quad\quad\left.+\frac 1{48} R (S^{(2)})^3f^{(6)}(S^\prime)\right)\\
   +&\mathcal O(\hbar^4),
 \end{array}
\end{equation}
we find by separating the real and imaginary part of $e^{-\frac i \hbar S}$ times Schr\"odinger's equation that
\begin{equation}
 \begin{array}{rcl}
   \dot R &=& -\left(R^\prime f^\prime(S^\prime)+\frac 1 2 r S^{(2)}f^{(2)}(S^\prime)\right)\\
          & &+\hbar^2\left(\left(\frac{1}{4}R^{(2)}S^{(2)}+\frac1{4!}RS^{(4)}\right)f^{(4)}(S^\prime)+\left(\frac 1 8 R^\prime (S^{(2)})^2+\frac 1 {12} R S^{(3)}S^{(2)}\right)f^{(5)}(S^\prime)\right.\\
          & &\quad\quad\left.+\frac{1}{48} R\,(S^{(2)})^3 f^{(6)}(S^\prime)\right)\\
          & &\,+\,\,\mathcal O(\hbar^4)\\
   \dot S &=& -\left(f(S^\prime)+g\right)\\
          & &+\hbar^2\left(\frac 1 2 \frac{R^{(2)}}{R} f^{(2)}(S^\prime)+\left(\frac 1 2 \frac{R^\prime S^{(2)}}{R}+\frac 1 6 S^{(3)}\right) f^{(3)}(S^\prime)+\frac 1 8 (S^{(2)})^2 f^{(4)}(S^\prime)\right)\\
          & &\,+\,\,\mathcal O(\hbar^4).

 \end{array}
\end{equation}
The leading order in the $\hbar$-expansion for $S$ is the classical Hamilton-Jacobi equation for a classical Hamiltonian $H^{cl}=f(p)+g(q)$ and the leading order quantum corrections to this Hamiltonian are
\begin{equation}\label{equ:nonStandardHJ}
 \delta\,H=-\hbar^2\left.\left(\frac{1}{2}\frac{R^{(2)}}{R} H^{cl}_{,pp}+\frac 1 2 \left( \frac{R^\prime S^{(2)}}{R}+\frac 1 3 S^{(3)}\right) H^{cl}_{,ppp}+\frac 1 8 (S^{(2)})^2 H^{cl}_{,pppp}\right)\right|_{p=S^\prime(q)}
\end{equation}
The next step consists of finding an expression for the probability current $\vec{j}_\psi$. This is ambiguous in more than one dimension, since equivariance with the quantum equilibrium condition determines only the divergence of the current, so one is free to add an arbitrary divergence-free vector field to the current. This ambiguity can be fixed in part by requiring that the $\hbar$-expansion of the velocity field $\vec{v}:=\vec{j_\psi}/R^2$ is semiclassical, i.e. $\vec v=\left.\{q,H^{cl}\}\right|_{p=S^\prime}+\mathcal O(\hbar^2)$. A current with this property has been proposed by Struyve and Valentini \cite{Struyve:2008va} for arbitrary dimensions, which simplifies for dimensional Hamiltonian $\hat H=\sum_{n=0}^\infty h_n(q) \partial^n$ to:
\begin{equation}
 j_\psi:=\frac{i}{\hbar}\sum_{n=1}^\infty \sum_{m=0}^{n-1}(-1)^m (\psi^* h_n^*)^{(m)}\,\psi^{(n-m-1)}.
\end{equation}
Inserting a Hamiltonian of the form (\ref{equ:nonStandardHamiltonianForm}) into this current formula and performing an $\hbar$ expansion of the associated velocity field, one finds
\begin{equation}\label{equ:nonStandardVelocity}
 \begin{array}{l}
   v=\left.\{q,H^{cl}\}\right|_{p=S^\prime}-\hbar^2\int^q\left[\left(\frac{R^{(2)}S^{(2)}}{2R}+\frac{S^{(4)}}{12} \right)H^{cl}_{,pppp}+\frac{(S^{(2)})^3}{24}H_{,pppppp}\right.\\ 
  \quad\quad\quad\quad\quad\quad\quad\quad\quad\quad\quad\quad\left.\left.+\left(\frac{R^\prime(S^{(2)})^2}{4\,R}+\frac{S^{(2)}S^{(3)}}{6}\right)H^{cl}_{,ppppp}
  \right]\right|_{p=S^\prime} +\mathcal O(\hbar^4)
 \end{array}
\end{equation}
Following the reasoning of the previous section, the $\hbar^2$-orders of equation (\ref{equ:nonStandardHJ}) and (\ref{equ:nonStandardVelocity}) are the leading order quantum corrections to classical evolution. 

\section{Effective Equations of Motion}\label{appendix:EffectiveEOM}

A velocity field $v=\{q,H^{cl}\}|_{p=S^\prime}+\hbar^2 v_{qu}+\mathcal O(\hbar^3)$ shows immediately that the evolution of the configuration $q$ is well approximated by the classical equations of motion $\dot q=\{q,H^{cl}\}$ if the $\mathcal O(\hbar^2)$ contributions are negligible. However, given and effective Hamilton-Jacobi equation $\dot S+H^{cl}|_{p=S^\prime}+\hbar^2 Q+\mathcal O(\hbar^3)=0$ it does not immediately follow that the evolution of the effective momentum $p=S^\prime(q)$ is well approximated by the classical equations of motion if the $\mathcal O(\hbar^2)$ contributions to the effective Hamilton-Jacobi equation are negligible. Rather we have to use the effective Hamilton-Jacobi equation to derive the the effective equations of motion. 

A straightforward way to derive the effective equations of motion is to insert the ansatz $S(x)=\sum_{n=0}^\infty s_n (x-q)^n$, where $s_1=p$, and $R(x)=\sum_{n=0}^\infty r_n (x-q)^n$, where $r_o>0$, into the effective Hamilton-Jacobi equation and to extract the coefficient of $(x-q)^1$. The leading order in the $\hbar$-expansion of this coefficient is $\dot s_1 = 2 s_2 H^{cl}_{,p}(s_1,q)-\left(H^{cl}_{,q}(s_1,q)+2s_2 H^{cl}_{,p}(s_1,q)\right)+\mathcal O(\hbar^2)$, which, upon the replacement $s_1 \to p$, reduces to the classical equations of motion
\begin{equation}
 \dot p = - H^{cl}_{,q}(p,q)+\mathcal O(\hbar^2).
\end{equation}
The order $\hbar^2$ of the expansion gives the leading quantum corrections to these classical equations of motion:
\begin{equation}\label{equ:dotPsubleading}
 \dot p = - H^{cl}_{,q}(p,q)+\hbar^2\left(2 s_2\,v_{qu}(q)-\left.Q_{,x}\right|_{x=q,s_1=p}\right) +\mathcal O(\hbar^3),
\end{equation}
so we expect the dynamics of the system to be well approximated by the classical equations of motion when the leading quantum corrections to the classical equations motion are small, i.e. if:
\begin{equation}
  \hbar^2 v_{qu}(q) \ll H^{cl}_{,p}\,\quad\,\textrm{and }\,\quad\,\hbar^2\left(\left.Q_{,x}\right|_{x=q,s_1=p}-2 s_2\,v_{qu}(q)\right)\ll H^{cl}_{,q}. 
\end{equation}

\end{appendix}


\begin{thebibliography}{xxx}
 \bibitem{Anderson:2004bq} E.~Anderson, J.~Barbour, B.~Z.~Foster, B.~Kelleher and N.~O.~Murchadha: ``The Physical gravitational degrees of freedom,'' Class.\ Quant.\ Grav.\  {\bf 22} (2005) 1795 [gr-qc/0407104]
 \bibitem{Barbour:2011dn} J.~Barbour: ``Shape Dynamics: An Introduction,'' arXiv:1105.0183 [gr-qc]
 \bibitem{Barbour:2010dp} J.~Barbour: ``The Definition of Mach's Principle,'' Found.\ Phys.\  {\bf 40} (2010) 1263 [arXiv:1007.3368 [gr-qc]]
 \bibitem{Gomes:2010fh} H.~Gomes, S.~Gryb and T.~Koslowski: ``Einstein gravity as a 3D conformally invariant theory,'' Class.\ Quant.\ Grav.\  {\bf 28} (2011) 045005 [arXiv:1010.2481 [gr-qc]]
 \bibitem{Gomes:2011zi} H.~Gomes and T.~Koslowski: ``The Link between General Relativity and Shape Dynamics,'' Class.\ Quant.\ Grav.\  {\bf 29} (2012) 075009 [arXiv:1101.5974 [gr-qc]]
 \bibitem{Gomes:2011au} H.~Gomes and T.~Koslowski: ``Coupling Shape Dynamics to Matter Gives Spacetime,'' Gen.\ Rel.\ Grav.\  {\bf 44} (2012) 1539 [arXiv:1110.3837 [gr-qc]]
 \bibitem{Gomes:2012hq} H.~Gomes and T.~Koslowski: ``Frequently asked questions about Shape Dynamics,'' Found.\ Phys.\  {\bf 43} (2013) 1428 [arXiv:1211.5878 [gr-qc]]
 \bibitem{Koslowski:2012uk} T.~A.~Koslowski: ``Observable Equivalence between General Relativity and Shape Dynamics,'' arXiv:1203.6688 [gr-qc]
 \bibitem{Koslowski:2013gua} T.~A.~Koslowski: ``Shape Dynamics and Effective Field Theory,'' Int.\ J.\ Mod.\ Phys.\ A {\bf 28} (2013) 1330017 [arXiv:1305.1487 [gr-qc]]
 \bibitem{Bojowald:2008zzb} M.~Bojowald: ``Loop quantum cosmology,'' Living Rev.\ Rel.\  {\bf 11} (2008) 4
 \bibitem{Arnowitt:1962hi} R.~L.~Arnowitt, S.~Deser and C.~W.~Misner: ``The Dynamics of general relativity,'' Gen.\ Rel.\ Grav.\  {\bf 40} (2008) 1997 [gr-qc/0405109]
 \bibitem{Dirac:1958ha} P.A.M. Dirac: ``The Theory of Gravitation in Hamiltonian Form,'' Proc. Roy. Soc. London A (1958) 246 (1246) 333–343
 \bibitem{Dirac:1959fi} P. A. M. Dirac: ``Fixation of coordinates in the Hamiltonian theory of gravitation,'' Phys. Rev. 114 (1959) 924–930
 \bibitem{York:1972ro} J. W. York: ``Role of conformal three geometry in the dynamics of gravitation,'' Phys. Rev. Lett. 28 (1972) 1082–1085
 \bibitem{York:1973cm} J.~W.~York: ``Conformally invariant orthogonal decomposition of symmetric tensors on Riemannian manifolds and the initial value problem of general relativity,'' J.\ Math.\ Phys.\  {\bf 14} (1973) 456
 \bibitem{Bohm:1952aa} D. Bohm: ``A Suggested Interpretation in Terms of 'Hidden Variables': Part I and Part II'' , Phys. Rev. 85, 166–179 and 180–193 (1952)
 \bibitem{Allori:2002dg} V. Allori, D. D\"urr, S. Goldstein and N. Zanghi: ``Seven steps towards the classical world,'' J. Opt. B: Quantum Semiclass. Opt. (2002) 4 S482 [arXiv:quant-ph/0112005]
 \bibitem{Anderson:2010xm} E.~Anderson: ``The Problem of Time in Quantum Gravity,'' arXiv:1009.2157 [gr-qc]
 \bibitem{Barbour:1982gha} J.~B.~Barbour and B.~Bertotti: ``Mach's Principle and the Structure of Dynamical Theories,'' Proc.\ Roy.\ Soc.\ Lond.\ A {\bf 382} (1982) 295
 \bibitem{Barbour:2013qb} J.~Barbour, M.~Lostaglio and F.~Mercati: ``Scale Anomaly as the Origin of Time,'' Gen.\ Rel.\ Grav.\  {\bf 45} (2013) 911 [arXiv:1301.6173 [gr-qc]]
 \bibitem{Barbour:2013goa} J.~Barbour, T.~Koslowski and F.~Mercati: ``The Solution to the Problem of Time in Shape Dynamics,'' arXiv:1302.6264 [gr-qc]
 \bibitem{Barbour:2013jya} J.~Barbour, T.~Koslowski and F.~Mercati: ``A Gravitational Origin of the Arrows of Time,'' arXiv:1310.5167 [gr-qc]
 \bibitem{Rovelli:2004tv} C.~Rovelli: ``Quantum gravity,'' Cambridge, UK: Univ. Pr. (2004)
 \bibitem{Bojowald:2010xp}  M.~Bojowald, P.~Hoehn and A.~Tsobanjan: ``An Effective approach to the problem of time,'' Class.\ Quant.\ Grav.\  {\bf 28} (2011) 035006 [arXiv:1009.5953 [gr-qc]]
 \bibitem{Moncrief:1989dx} V.~Moncrief: ``Reduction of the Einstein equations in (2+1)-dimensions to a Hamiltonian system over Teichmuller space,''
  J.\ Math.\ Phys.\  {\bf 30} (1989) 2907
 \bibitem{Budd:2011er} T.~Budd and T.~Koslowski: ``Shape Dynamics in 2+1 Dimensions,'' Gen.\ Rel.\ Grav.\  {\bf 44} (2012) 1615 [arXiv:1107.1287 [gr-qc]]
 \bibitem{Struyve:2008va} W. Struyve and A. Valentini: ``De Broglie-Bohm Guidance Equations for Arbitrary Hamiltonians,'' J. Phys. A 42, 035301 (2009) [arXiv:0808.0290]
\end{thebibliography}
\end{document}